%% file: paper.tex
\documentclass[smallextended]{orsvjour3}

\usepackage{amssymb,hyperref,color,graphicx}
\usepackage{mathrsfs}
\DeclareSymbolFontAlphabet{\mathrsfs}{rsfs}

\newcommand{\tfrac}[2]{\textstyle \frac{#1}{#2}}
\newcommand{\half}{\tfrac{1}{2}}
\newcommand{\diag}{\mathrm{diag}}
\newcommand{\const}{\mathrm{const}}
\newcommand{\Krr}{K^r{}_r}
\newcommand{\hKrr}{{\hat K^r{}_r}}
\newcommand{\hxi}{{\hat \xi}}
\newcommand{\hPi}{{\hat \Pi}}
\newcommand{\eref}[1]{(\ref{#1})}
\newcommand{\rmd}{\mathrm{d}}
\newcommand{\rme}{\mathrm{e}}
\newcommand{\etal}{\textit{et al.\,\,}}
 
\begin{document}

\title{Type II critical collapse on a single fixed grid:\\
  a gauge-driven ingoing boundary method}
\titlerunning{Type II critical collapse on a single fixed grid}
\author{Oliver Rinne} 
\institute{Hochschule f\"ur Technik und Wirtschaft Berlin,
Treskowallee 8, 10318 Berlin, Germany 
\and \\
Max Planck Institute for Gravitational Physics
(Albert Einstein Institute), Am M\"uhlenberg 1, 14476 Potsdam, Germany 
\\
\email{oliver.rinne@htw-berlin.de}
}

\maketitle

\begin{abstract}
  We develop a numerical method suitable for gravitational collapse based on Cauchy 
  evolution with an ingoing characteristic boundary.
  Unlike similar methods proposed recently (Ripley; Bieri, Garfinkle \& Yau 2019/20),
  the numerical grid remains fixed during the evolution and no points need to be removed
  or added. 
  Increasing coordinate refinement of the central region as the field collapses is achieved
  solely through the choice of spatial gauge and particularly its boundary condition.
  We apply this method to study critical collapse of a massless scalar field in spherical
  symmetry using maximal slicing and isotropic coordinates.
  Known results on mass scaling, discrete self-similarity and universality of the
  critical solution (Choptuik 1993) are reproduced using this
  considerably simpler numerical method.

  \keywords{Numerical relativity \and Boundary conditions \and 
    Black holes\and Critical collapse}  
  %\PACS{04.20.-q %classical gr 
  %      \and
  %      04.25.D- %num rel 
  %}
\end{abstract}

%%%%%%%%%%%%%%%%%%%%%%%%%%%%%%%%%%%%%%%%%%%%%%%%%%%%%%%%%%%%%%%%%%%%%%%%%%%%%%%

\section{Introduction}
\label{s:intro}

Critical phenomena in gravitational collapse are one of the most remarkable discoveries 
made through numerical methods applied to Einstein's field equations of 
general relativity.
Since Choptuik's groundbreaking study of a massless scalar field coupled to the Einstein 
equations in spherical symmetry \cite{Choptuik1993}, similar phenomena have been 
discovered for a variety of matter models and even in vacuum, see \cite{Gundlach2007}
for a review article.
Briefly, the idea is to choose a smooth one-parameter family of initial data 
such that in the future Cauchy development of such data, a black hole forms for large
parameter values and the field disperses to flat spacetime for small parameter values.
We are interested in the threshold between these two final states and the associated critical solution.
In what has been termed Type II critical collapse, the black hole mass becomes
infinitesimally small as the threshold, obeying a universal scaling law, and the critical solution is discretely self-similar and universal, i.e. independent of 
the particular one-parameter family of initial data chosen.
(There is also Type I critical collapse in certain models, where the black hole mass is finite at the threshold and the critical solution is stationary or time-periodic.)

It is this discrete self-similarity of near-critical evolutions that makes the problem
so hard numerically: the solution repeats itself on smaller and smaller spatial scales,
in shorter and shorter time intervals.
Choptuik \cite{Choptuik1993} implemented an adaptive mesh refinement algorithm
\cite{Berger1984} in order to be able to resolve the increasingly smaller length scales of the solution.

Alternate methods to tackle the same problems have subsequently been developed,
e.g. formulations in double null coordinates with \cite{Hamade1996} and without 
\cite{Garfinkle1995} adaptive mesh refinement, although in the latter case grid points
had to be added during the evolution in order to maintain accuracy.

Recently Bieri, Garfinkle and Yau \cite{Bieri2020} proposed a general method for Cauchy 
evolution in numerical relativity whereby the boundary of the finite spatial computational 
domain is expanded along a spacelike direction at each time step.
Additional initial data must be specified on this surface.
The advantage is that with such a setup, no outer boundary conditions need to be imposed
because all the constant-time slices lie within the domain of dependence of the initial
slice and the additional ``tilted'' spacelike surface.
This proposal thus avoids the long-standing problem of imposing boundary conditions along
a finite timelike surface in general relativity \cite{Sarbach2007}. 
Other alternatives to this problem include Cauchy-characteristic matching 
\cite{Winicour2012}, evolution on hyperboloidal slices compactified towards future 
null infinity \cite{Zenginoglu2008,Moncrief2009,Rinne2010} 
and the regular conformal field equations \cite{Friedrich1983,Frauendiener2004}.

Related to Bieri \etal's scheme is the ``excision method'' proposed by
Ripley \cite{Ripley2019}, whereby the computational domain is excised along a surface that 
is spacelike or tangent to an ingoing characteristic of the boundary on the initial slice.
Again, no boundary conditions need to be impose at the outer boundary because in this case 
all characteristics leave the computational domain.
This method appears to be well suited to gravitational collapse problems.
A disadvantage is that grid points are lost during the evolution due to the excision 
procedure so that one will have to add grid points in the interior in order to maintain 
accuracy.

The method developed in the present paper is similar to Ripley's in that the outer
boundary of the spatial computational domain is an ingoing characteristic.
However, no grid points are excised; instead merely the spatial coordinates are
changed as time proceeds.

In the Arnowitt-Deser-Misner (ADM) formulation of general 
relativity \cite{Arnowitt1962}, the time vector field $\partial/\partial t$ is 
decomposed as\footnote{
  We use abstract index notation with indices $a,b,\ldots$ ranging over the spacetime 
  coordinates $t,r,\theta,\varphi$ and indices $i,j,\ldots$ ranging over the spatial
  coordinates $r,\theta,\varphi$.
}
\begin{equation} 
    \left(\tfrac{\partial}{\partial t}\right)^a = \alpha n^a + \beta^a, 
\end{equation}
where $n_a = -\alpha \nabla_a t$ is the unit future-directed timelike normal, 
$\alpha$ is the lapse function and $\beta^a$ the shift vector (which is spatial,
$n_a \beta^a = 0$).
This means that a point with spatial coordinates $x^i$ on the spatial slice at 
time $t$ correponds to the point with spatial coordinates 
$x^i - \beta^i \rmd t$ on the slice at time $t + \rmd t$ if we drag it along the timelike normal.

Consider now a shift vector field of the form 
\begin{equation} \label{e:isoshift}
  \beta^i = c x^i
\end{equation}  
where $c$ is a constant w.r.t. the spatial coordinates $x^i$.
Hence identified points on the spatial slices will change coordinates according to 
\begin{equation} \label{e:coordchange}
    x^i \rightarrow (1 - c \, \rmd t) x^i  
\end{equation}
as time increases from $t$ to $t+\rmd t$, so if we choose $c<0$ then the coordinates 
``zoom in" isotropically towards the origin.
The significance of \eref{e:isoshift} is that it is a homogeneous solution to the
spatial isotropic gauge condition in spherical symmetry, Eq. \eref{e:isotropic} below,
where $\beta^r = r \beta$ so \eref{e:isoshift} corresponds to $\beta=c=\const$.
The value of the constant $c$ will be fixed by the boundary condition on the shift
in the isotropic gauge condition.
For a suitable value of this (in general time-dependent) constant, the outer boundary 
can be made an ingoing characteristic (or spacelike) so that no boundary conditions on 
the evolved fields are needed.

We supplement the isotropic spatial gauge condition with a maximal slicing condition.
The advantage of such a slicing as compared with the polar slicing used by Choptuik 
\cite{Choptuik1993} is that the coordinates remain regular at the apparent horizon
when it forms, which allows for a more accurate determination of its location and mass.

This article is organised as follows.
In Sect. \ref{s:form} we set up our model problem of a massless scalar field in 
spherical symmetry, and we state the gauge conditions and their boundary conditions
appropriate for our ingoing boundary method. 
In Sect. \ref{s:num} we provide details on the numerical methods we use to solve
the field equations.
The numerical results are contained in Sect. \ref{s:results}.
We set up two families of initial data, describe our method to tune to the critical
parameter and to choose an appropriate outer boundary radius, and we present results 
on the mass scaling, discrete self-similarity and universality of the critical solution.
In Sect. \ref{s:disc} we summarise, discuss potential challenges of the method and
future applications.

%%%%%%%%%%%%%%%%%%%%%%%%%%%%%%%%%%%%%%%%%%%%%%%%%%%%%%%%%%%%%%%%%%%%%%%%%%%%%%%
%%%%%%%%%%%%%%%%%%%%%%%%%%%%%%%%%%%%%%%%%%%%%%%%%%%%%%%%%%%%%%%%%%%%%%%%%%%%%%%

\section{Formulation of the model} 
\label{s:form}

\subsection{Choice of gauge and variables}

In spherical symmetry and isotropic coordinates, the spacetime metric takes the form 
\begin{equation} \label{e:metric} 
  ds^2 = -(\alpha^2 - \psi^4 r^2 \beta^2) \rmd t^2 + 2 r \beta \psi^4 \rmd t \, \rmd r
    + \psi^4 \left[\rmd r^2 + r^2(\rmd\theta^2 + \sin^2\theta \, \rmd\varphi^2)\right]. 
\end{equation}
We impose maximal slicing and hence the extrinsic curvature has only one independent component 
in spherical symmetry:
\begin{equation}
  K^i{}_j = \diag(\Krr, -\half \Krr, -\half \Krr).
\end{equation}  
For reasons discussed shortly, we define a rescaled quantity
\begin{equation} \label{e:hKrrdef}
  \hKrr := r^{-2} \psi^6 \Krr.
\end{equation}
The massless scalar field $\Phi$ itself does not enter the equations but only its first 
derivatives
\begin{equation} \label{e:hxihPidef}
  \hxi := r^{-1} \psi^2 \Phi', \qquad 
  \hPi := \psi^4 \alpha^{-1}(\dot \Phi - r\beta \Phi'),
\end{equation}  
where here and in the following a dash denotes a partial derivative w.r.t. $r$ and a dot 
w.r.t. $t$.

The fundamental variables $\alpha, \beta, \psi, \hKrr, \hxi$ and $\hPi$ depend on $t$ and $r$ only, 
and the powers of $r$ in their definitions have been chosen so that they are all 
even functions of $r$ with finite nonzero limits at the origin $r=0$.

%%%%%%%%%%%%%%%%%%%%%%%%%%%%%%%%%%%%%%%%%%%%%%%%%%%%%%%%%%%%%%%%%%%%%%%%%%%%%%%

\subsection{Field equations}

The relevant components of the Einstein equations $R_{ab} = \kappa \nabla_a \nabla_b \Phi$, 
where $R_{ab}$ is the spacetime Ricci tensor and $\kappa=8\pi$ in geometric units, are the
momentum constraint
\begin{equation} \label{e:momcons}
  r \hKrr' + 5 \hKrr + \kappa \hat\pi \hat\xi = 0
\end{equation}  
and the Hamiltonian constraint
\begin{equation} \label{e:hamcons}
  \psi'' + 2 r^{-1} \psi' + \tfrac{3}{16} \psi^{-7} (r^2 \hKrr)^2
  + \tfrac{1}{8} \kappa \psi^{-3} (\hPi^2 + r^2\hxi^2) = 0.
\end{equation}  
When linearising \eref{e:hamcons} about a given background solution 
$\overline{\psi}$, the coefficient of the undifferentiated term proportional to 
$\psi$ is manifestly negative because of the negative powers of $\psi$ in 
\eref{e:hamcons}.
If this was not the case then non-unique oscillatory solutions might exist; see 
\cite{Walsh2007,Rinne2008} for further discussion of this issue.
This is the reason for the choice of the powers of $\psi$ in \eref{e:hKrrdef} and 
\eref{e:hxihPidef}.

The maximal slicing condition implies the following equation for the lapse,
\begin{equation} \label{e:slicing}
  \alpha'' + 2 \alpha' (r^{-1} + \psi^{-1} \psi') - \alpha\left[
    \kappa \psi^{-4} \hPi^2 + \tfrac{3}{2} \psi^{-8} (r^2\hKrr)^2\right] = 0,
\end{equation}  
and preservation of the isotropic form of the metric \eref{e:metric} yields 
\begin{equation} \label{e:isotropic}
  \beta' - \tfrac{3}{2} r\alpha\psi^{-6}\hKrr = 0.
\end{equation}  

The equation of motion for the scalar field $\nabla^a \nabla_a \Phi = 0$ 
reduces to the pair of first-order equations
\begin{eqnarray}
  \label{e:dothxi}
  \dot \hxi &=& r\beta\hxi' + (3\beta + 2\alpha\psi^{-6}r^2\hKrr)\hxi
    + \alpha\psi^{-2}r^{-1}\hPi' \nonumber\\
    &&+ \psi^{-3} r^{-1} (\psi\alpha' - 4\alpha\psi')\hPi, \\
  \label{e:dothPi}  
  \dot \hPi &=& r\beta\hPi' + (2\beta + \alpha\psi^{-6}r^2 \hKrr)\hPi
    + \alpha\psi^{-2}r\hxi' + \psi^{-2}(r\alpha' + 3\alpha)\hxi.  
\end{eqnarray}

There are redundant evolution equations for $\psi$ and $\hKrr$ that can be used to
monitor the accuracy of the code; the first will also be needed to specify boundary conditions:
\begin{eqnarray} 
  \label{e:dotpsi}
  \dot\psi &=& r\beta\psi' + \half\beta\psi 
    + \tfrac{1}{4}r^2 \alpha\psi^{-5}\hKrr, \\
  \label{e:dothKrr}
  \dot \hKrr &=& r\beta \hKrr' + 5\beta\hKrr 
    + \tfrac{3}{2}\alpha r^2\psi^{-6}\hKrr^2 
    - \tfrac{2}{3}\psi^2 r^{-1} (r^{-1} \alpha')' \nonumber\\
    &&- \tfrac{4}{3} \alpha\psi r^{-1}(r^{-1}\psi')'
    + 4r^{-2} \psi'(\alpha\psi' + \tfrac{2}{3}\psi\alpha')
    - \tfrac{2}{3}\kappa\alpha\psi^{-2}\hxi^2. 
\end{eqnarray}  

\subsection{Boundary conditions}

A crucial feature of our method is the choice of gauge boundary conditions.
We want the outer boundary to be ingoing null or spacelike, which corresponds to 
setting
\begin{equation} \label{e:bcbeta}
  \beta \doteq -\nu \, r^{-1} \psi^{-2} \alpha
\end{equation}  
with $\nu\geqslant 1$, where $\doteq$ means equality at the outer boundary 
$r=r_{\max}$.
For the results presented in Sect. \ref{s:results} we will always choose $\nu = 1$ 
corresponding to the boundary being null, although we will briefly discuss 
making $\nu$ a time-dependent function in Sect. \ref{s:disc}.

Since there are no ingoing characteristics at the outer boundary with this choice, 
the evolution equations \eref{e:dothxi} and \eref{e:dothPi} for the scalar field
do not require any boundary conditions.
We specify Dirichlet boundary conditions on $\psi$ for the Hamiltonian constraint
\eref{e:hamcons} by evolving \eref{e:dotpsi} at the outer boundary.
The momentum constraint \eref{e:momcons} does not require a boundary condition 
as this is already fixed by demanding the solution to be regular at the origin.

What remains to be specified is an outer boundary condition on the lapse $\alpha$ for
the maximal slicing condition \eref{e:slicing}.
Freezing the lapse to its flat value $\alpha=1$ is not a good idea since this will 
lead to unacceptably large slice stretching as the physical size of the grid shrinks 
and the lapse collapses in the centre as the singularity is approached.
Instead we simply advect the lapse along the shift at the outer boundary, as in the first terms 
of all the evolution equations:
\begin{equation} \label{e:dotalpha}
  \dot \alpha \doteq r\beta\alpha'.
\end{equation} 
Another way of phrasing this is to extrapolate (in time) the value of the lapse at the outer 
boundary $r=r_{\max}$ on the slice at time $t+\rmd t$ from its value at the identified radius
on the slice at time $t$, which according to \eref{e:coordchange} is at 
$(r + r\beta \rmd t)_{r=r_{\max}} < r_{\max}$ (note $\beta < 0$ at $r=r_{\max}$).

%%%%%%%%%%%%%%%%%%%%%%%%%%%%%%%%%%%%%%%%%%%%%%%%%%%%%%%%%%%%%%%%%%%%%%%%%%%%%%%
%%%%%%%%%%%%%%%%%%%%%%%%%%%%%%%%%%%%%%%%%%%%%%%%%%%%%%%%%%%%%%%%%%%%%%%%%%%%%%%

\section{Numerical methods}
\label{s:num}

\subsection{Evolution scheme}

Given data at time $t$, we first evolve the scalar field variables $\hxi$ and $\hPi$
to the next timestep $t+\Delta t$ using \eref{e:dothxi} and \eref{e:dothPi}.
At the advanced time the radial ordinary differential equations (ODEs) 
\eref{e:momcons}--\eref{e:isotropic} are solved
in this order for $\hKrr, \psi, \alpha$ and $\beta$ (notice they form a hierarchy).
Dirichlet boundary values for $\psi$ and $\alpha$ are supplied by evolving
\eref{e:dotpsi} and \eref{e:dotalpha} at the outer boundary along with the other 
evolution equations, and the boundary condition for $\beta$ is \eref{e:bcbeta}.

%%%%%%%%%%%%%%%%%%%%%%%%%%%%%%%%%%%%%%%%%%%%%%%%%%%%%%%%%%%%%%%%%%%%%%%%%%%%%%%

\subsection{Discretisation}

We use a fixed non-uniform radial grid at points $r_i = f(x_i)$, where
\begin{equation} 
  f:[0,1]\to[0,r_{\max}], \quad f(x) = 
    r_\mathrm{eff} x + (r_{\max} - r_\mathrm{eff}) x^3
\end{equation}
is a cubic map from numerical to physical coordinates.
Here $r_\mathrm{eff} \leqslant r_{\max}$ can be thought of as an ``effective'' radius the grid 
would have if the same resolution as close to the origin was used all the way to the
outer boundary.
We typically choose $r_\mathrm{eff} \approx \half r_{\max}$.
It should be noted that a non-uniform grid is not essential for our method to work, 
it just saves computational resources since the distribution of grid points is better adapted to
the features of the solution, which has its largest gradients close to the origin.  
We could just as well take $r_\mathrm{eff} = r_{\max}$ corresponding to a uniform 
grid.
With respect to the numerical coordinate $x$, the grid is equidistant and staggered at the 
origin: 
\[ x_i = (i + \half) h, \quad  i = 0, 1, \ldots, N, \quad  h = (N + \half)^{-1}. \]
We use $N=500$ grid points for the simulations presented in Sect. \ref{s:results}.
The grid remains unchanged during the evolution.

We use centred fourth-order finite differences to discretise the equations in $r$.
Near the origin the finite-difference stencils are modified according to the known (even) 
$r$-parity of all the evolved variables.
Near the outer boundary (fourth-order) backward finite differences are used.

%%%%%%%%%%%%%%%%%%%%%%%%%%%%%%%%%%%%%%%%%%%%%%%%%%%%%%%%%%%%%%%%%%%%%%%%%%%%%%%

\subsection{ODE solvers}

Following the method of lines, the evolution equations are integrated forward in time 
using a standard fourth-order Runge-Kutta method.
Sixth-order Kreiss-Oliger dissipation \cite{Kreiss1973} is added to the right-hand 
sides of the evolution equations in order to maintain numerical stability (a small 
coefficient $\approx 0.1$ is found to be sufficient).

The radial ODEs are solved using a direct band-diagonal solver at each substep of the
Runge-Kutta method.

Since the size of the metric functions $\alpha,\beta$ and $\psi$ changes drastically 
during the evolution, it is important to adapt the size of the time step $\Delta t$
in order not to violate the Courant-Friedrichs-Lewy (CFL) condition for 
numerical stability.
At each time step, we compute the characteristic speeds of the scalar wave equation
\begin{equation}
  v_\pm(r) = -r\beta \pm \psi^{-2}\alpha
\end{equation}
and set the time step size according to 
\begin{equation}
  \Delta t = \lambda \min_{i=1,\ldots,N} 
    \frac{r_i - r_{i-1}}{\max(|v_+(r_i)|,|v_-(r_i)|)}.
\end{equation}  
The CFL condition states $0 < \lambda < 1$, and we typically choose $\lambda=\half$.

%%%%%%%%%%%%%%%%%%%%%%%%%%%%%%%%%%%%%%%%%%%%%%%%%%%%%%%%%%%%%%%%%%%%%%%%%%%%%%%

\subsection{Termination criteria}

We terminate a simulation when either a black hole forms (i.e. the evolution is 
\emph{supercritical}) or the field disperses to flat spacetime (i.e. the evolution is 
\emph{subcritical}).

Formation of a black hole is detected by looking for an apparent horizon 
(outermost marginally outer trapped surface).
This is an $r=\const$ surface whose outgoing null expansion vanishes,
\begin{equation} \label{e:ahcondgen}
  \theta_+ = 2 (\ln R)_{,a} \ell^a = 0,
\end{equation}  
where 
\begin{equation}
    R=r\psi^2
\end{equation}    
is the areal radius and $\ell^a$ is an outward-pointing radial null vector.
In our variables \eref{e:ahcondgen} is equivalent to
\begin{equation} \label{e:ahcond}
  r\psi' + \half\psi + \tfrac{1}{4} r^3 \psi^{-3}\hKrr = 0.
\end{equation}  
The radius $r_\mathrm{AH}$ of the apparent horizon is the largest zero of this 
equation, and the associated mass is
\begin{equation}
  M = \half R \vert_{r=r_\mathrm{AH}}.
\end{equation}
It is this mass computed from the apparent horizon that will enter the scaling law
in Sect. \ref{s:scaling}.
Assuming cosmic censorship holds, formation of an apparent horizon implies the 
existence of an event horizon containing the apparent horizon in its interior.

We consider an evolution to be subcritical if the maximum (w.r.t. $r$) of the scalar 
curvature 
\begin{equation} \label{e:Rscal}
  \mathcal{R} = \kappa\psi^{-8}\left[(r\hxi)^2 - \hPi^2\right]
\end{equation}
drops below some fraction (typically $5\%$) of its maximum value attained during the 
evolution.

%%%%%%%%%%%%%%%%%%%%%%%%%%%%%%%%%%%%%%%%%%%%%%%%%%%%%%%%%%%%%%%%%%%%%%%%%%%%%%%
%%%%%%%%%%%%%%%%%%%%%%%%%%%%%%%%%%%%%%%%%%%%%%%%%%%%%%%%%%%%%%%%%%%%%%%%%%%%%%%

\section{Numerical results}
\label{s:results}

\subsection{Initial data and bisection}

We consider two very different families of initial data for the scalar field:\\
(i) data that would be exactly \emph{ingoing} in a flat metric 
($\psi=\alpha=1, \, \beta=0$),
\begin{equation}
  \Phi = A \, \exp\left[ -\frac{1}{2} \left(\frac{r-r_0}{\sigma}\right)^2 \right],
  \quad \hxi = r^{-1} \Phi', \quad \hPi = r^{-1} (r\Phi)',
\end{equation}  
and (ii) data that are \emph{centred} at the origin and initially at rest,
\begin{equation}
  \Phi = A \, \exp\left[ -\frac{1}{2} \left(\frac{r}{\sigma}\right)^2 \right],
  \quad \hxi = r^{-1} \Phi', \quad \hPi = 0.
\end{equation}  
We fix the parameters $\sigma=1$ and (for the ingoing family) $r_0=10$, and we
take the amplitude $A$ as the critical parameter.
For large values of $A$ the solution forms a black hole whereas
for small values it disperses.
We use the bisection method to find an approximation to the critical amplitude $A_*$. 

%%%%%%%%%%%%%%%%%%%%%%%%%%%%%%%%%%%%%%%%%%%%%%%%%%%%%%%%%%%%%%%%%%%%%%%%%%%%%%%

\subsection{Choosing the outer boundary radius}

\begin{figure}
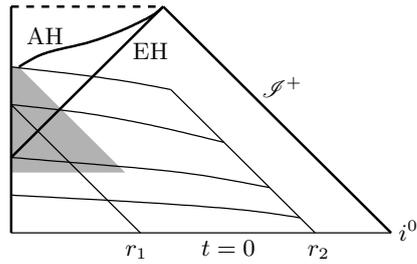
 
  \centering
  \input fig1_labels.tex 
  \caption{\label{f:penrose}
    Penrose diagram of a typical near-critical spacetime.
    Shown are the initial spatial slice at $t=0$ and a number of subsequent spatial
    slices and their ingoing null boundaries for two different initial boundary
    radii $r_1$ and $r_2$ as discussed in the main text.
    On the last slice the apparent horizon (AH) forms, which at later times
    converges to the event horizon (EH) of the black hole.
    Spacetime is close to the Type II critical solution roughly in the shaded region.
  }
\end{figure}

A typical Penrose diagram of a supercritical evolution close to the critical point
is shown in Fig. \ref{f:penrose}.
It becomes obvious from this diagram that the success of our method will depend 
on a good choice of the radius $r_{\max}$ of the outer boundary on the initial 
spatial slice.

If this is taken to be too large, $r_{\max} = r_2$ in Fig. \ref{f:penrose},
then despite the fact that the outer boundary is an ingoing characteristic,
the apparent horizon forms at a very small radius compared to the radius of the
outer boundary.
We terminate the bisection scheme if the radius of the apparent horizon in the supercritical
evolutions gets too small, say $r_\mathrm{AH} < 0.01 \, r_{\max}$, and start over 
with a smaller value of $r_{\max}$.

If on the other hand the initial boundary radius is chosen too small, 
$r_{\max} = r_1$ in Fig. \ref{f:penrose}, then the field escapes from the 
spatial domain before the apparent horizon forms.
In a numerical evolution of this type we observe that the bulk of the scalar field
moves out of the domain but the scalar curvature \eref{e:Rscal} remains large,
unlike in a subcritical evolution.
If this happens, we terminate the bisection scheme and repeat it with a larger value 
of $r_{\max}$.

Essentially this adds an outer bisection loop (for $r_{\max}$) to the inner one (for $A$).
In practice, one does not have to repeat the $A$-bisection all the way from the start
because one can use a somewhat smaller $A$-interval of the previous $r_{\max}$-iteration 
as the initial interval for the $A$-bisection at the improved value of $r_{\max}$.  

Using this procedure we determine $r_{\max}^{(i)} = 15.421875$ for the ingoing 
family and $r_{\max}^{(ii)} = 5.64$ for the centred family.
(For comparison, the near-critical ADM masses are $M_\mathrm{ADM}^{(i)} = 0.27$ and 
$M_\mathrm{ADM}^{(ii)} = 0.41$.)
Being able to observe the mass scaling (Sect. \ref{s:scaling}) does not require
such a precise choice of $r_{\max}$, while for the echoing behaviour of the critical
solution (Sect. \ref{s:echoing}) more accuracy is needed. (About three echos were 
visible in the simulations reported here.)

\begin{figure} 
  \centering
  \includegraphics[width=0.475\textwidth]{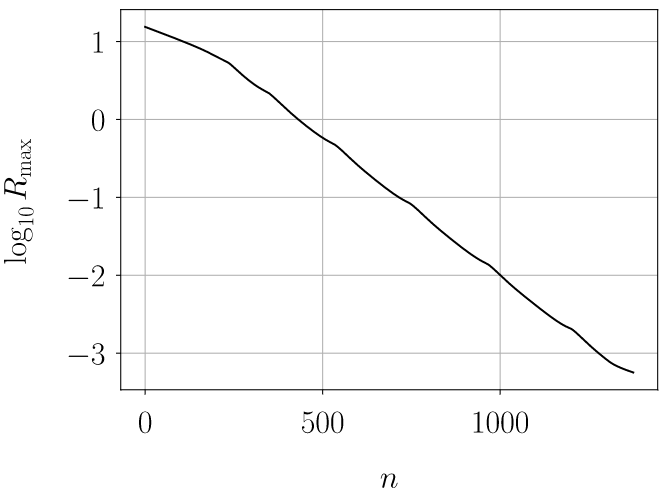}
  \includegraphics[width=0.475\textwidth]{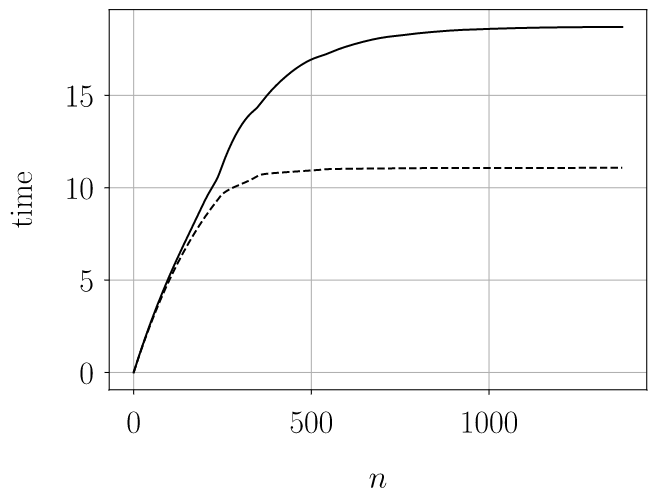}
  \caption{\label{f:gridevol}
    Left: areal radius $R_{\max}$ as a function of the number of time steps $n$.
    Right: coordinate time $t$ (solid curve) and proper time at the origin $T_0$ 
    (dashed curve) as functions of $n$.
  }
\end{figure}

To provide some idea of how the physical size of the grid changes during a simulation,
we plot in Fig. \ref{f:gridevol} the areal radius $R_{\max}$ of the outer boundary 
as a function of the number of time steps $n$ for a near-critical evolution.
The exponential decrease of $R_{\max}$ with $n$ is well adapted to the expected 
discrete self-similarity of the critical solution, which repeats itself on smaller
and smaller scales.
Also shown in Fig. \ref{f:gridevol} are the coordinate time $t$ and proper time 
at the origin 
\begin{equation}
  T_0 (t) = \int_0^t \alpha(\tilde t, 0) \rmd \tilde t
\end{equation}  
as functions of the number of time steps $n$. 
The latter approaches the accumulation time $T_0^*$ of the critical solution.

%%%%%%%%%%%%%%%%%%%%%%%%%%%%%%%%%%%%%%%%%%%%%%%%%%%%%%%%%%%%%%%%%%%%%%%%%%%%%%%

\subsection{Mass scaling}
\label{s:scaling}

\begin{figure} 
  \centering
  \includegraphics[width=0.55\textwidth]{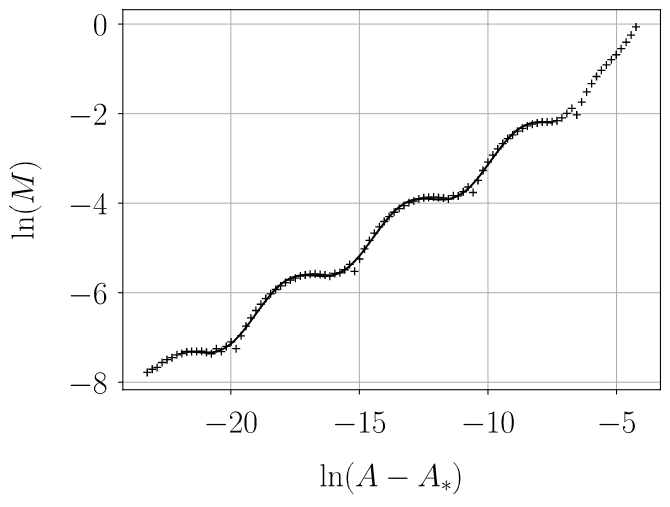}
  \caption{\label{f:scaling}
    Apparent horizon mass $M$ vs. critical parameter distance $A-A_*$
    in a double-logarithmic plot for a series of supercritical evolutions of the ingoing family ($+$) and best fit using the function
    $f(x) = c_0 + c_1 x + c_2 \cos (c_3 + c_4 x)$ (solid curve).
  } 
\end{figure}

In Fig. \ref{f:scaling} we plot the apparent horizon mass $M$ vs. the distance
$A - A_*$ to the critical amplitude for a series of supercritical evolutions.
In a double-logarithmic plot this forms a straight line with a periodic wiggle:
\begin{equation}
  \ln(M) = \gamma \ln(A - A_*) + \Psi[\ln(A - A_*)] + \const,
\end{equation} 
as first observed numerically in \cite{Choptuik1993,Hod1997} and predicted from a perturbative analysis of the critical solution in \cite{Hod1997,Gundlach1997}. 
According to this analysis, the period $\varpi$ of the function $\Psi$ is related to 
the echoing exponent $\Delta$ (cf. Sect. \ref{s:echoing}) via $\Delta = 2\varpi \gamma$. 
The values of the mass scaling exponent $\gamma$ and the echoing exponent $\Delta$ 
obtained from a fit to our numerical data are shown in Table \ref{t:scaling}
and are in good agreement with the predicted values.
It should be noted that $\Delta$ can be determined more accurately 
from the echoing behaviour of the near-critical solution 
(Sect. \ref{s:echoing}).

\begin{table}
  \begin{tabular}{l|l|l|l}
    & fit ingoing family (i) & fit centred family (ii) & prediction  \\ \hline
    $\gamma$ & $0.3744 \pm 0.0017$ & $0.3738 \pm 0.0027 $ & $0.374 \pm 0.001$ \\
    $\Delta$ & $3.419 \pm 0.033$ & $3.442 \pm 0.058$ & $3.4453 \pm 0.0005$
  \end{tabular}  
  \caption{\label{t:scaling}
    Mass scaling exponent $\gamma$ and echoing exponent $\Delta$ fitted from the 
    numerical values of the mass for the two initial data families, and their predictions
    from a perturbative analysis of the critical solution \cite{Gundlach1997}.
  }
\end{table}  

%%%%%%%%%%%%%%%%%%%%%%%%%%%%%%%%%%%%%%%%%%%%%%%%%%%%%%%%%%%%%%%%%%%%%%%%%%%%%%%

\subsection{Discrete self-similarity and universality of the critical solution}
\label{s:echoing}

Variables that are scale invariant display discrete self-similarity in near-critical
evolutions.
One such scale-invariant variable for the scalar field is 
\begin{equation} \label{e:Xdef}
    X := r \Phi' = r^2\psi^{-2}\hxi. 
\end{equation}    
Discrete self-similarity is best described in terms of logarithmic coordinates
\begin{equation}
  \tau := \ln (T_0^* - T_0), \qquad \rho := \ln R,
\end{equation}  
where $T_0^*$ is the accumulation time of the critical solution.
The conjecture, first discovered numerically in \cite{Choptuik1993}, is that
for the critical solution (indicated by the star), any scale-invariant variable 
such as $X$ \eref{e:Xdef} obeys
\begin{equation} \label{e:selfsim}
  X_*(\rho-\Delta, \tau-\Delta) = X_*(\rho,\tau),
\end{equation}
where $\Delta$ is the echoing exponent.

\begin{figure} 
  \centering
  \includegraphics[width=0.495\textwidth]{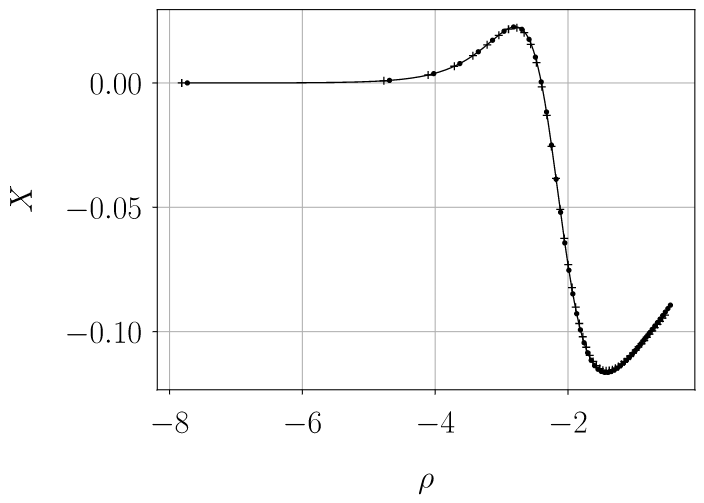}
  \includegraphics[width=0.495\textwidth]{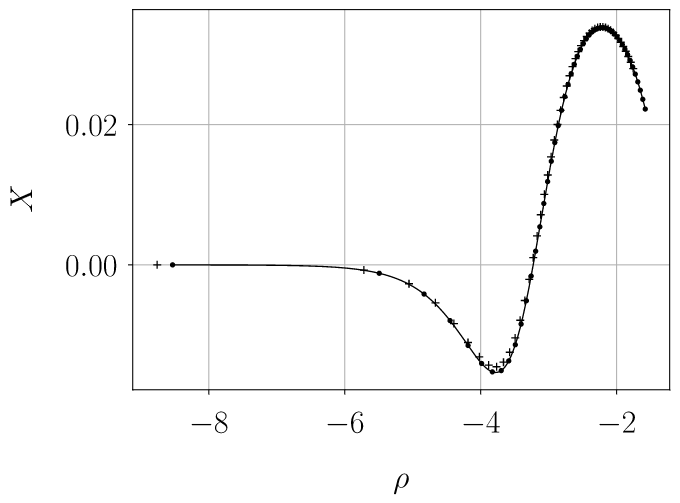}
  \caption{\label{f:echoing}
    Discrete self-similarity:
    the scale-invariant variable $X^{(i)}(\tau,\rho)$ for a near-critical evolution of 
    the ingoing family is plotted as a function of $\rho$
    at two different times $\tau = -1.8$ (left) and $\tau = -3.1$ (right) as a solid
    curve with dots at every tenth grid point.
    In the same plots, we also show $X^{(i)}(\tau-\Delta,\rho-\Delta)$ as a function of $\rho$ 
    with plus symbols ($+$) at every tenth grid point.
    The echoing exponent is taken to be $\Delta=3.44$.
  }
\end{figure}

In Fig. \ref{f:echoing} we plot $X^{(i)}(\rho,\tau)$ for a near-critical evolution of the
ingoing family as a function of $\rho$ at two different times $\tau$, and we overlay
$X^{(i)}(\rho-\Delta,\tau-\Delta)$ in the same plots, using $\Delta=3.44$.
The accumulation time $T_0^*$ has been determined by minimising the norm of the
difference between both functions at one fixed time $\tau$.
The fact that the curves nearly coincide provides strong support of the 
echoing property \eref{e:selfsim}.
We can also see in Fig. \ref{f:echoing} that the solution is well resolved numerically
both at the original time $\tau$ and at the time of the echo $\tau-\Delta$,
when the spatial scale has shrunk by a factor $\rme^\Delta \approx 31$.

\begin{figure} 
  \centering
  \includegraphics[width=0.495\textwidth]{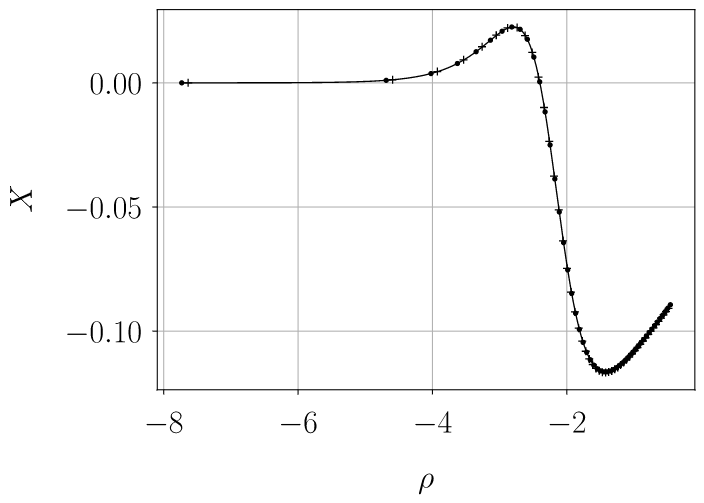}
  \includegraphics[width=0.495\textwidth]{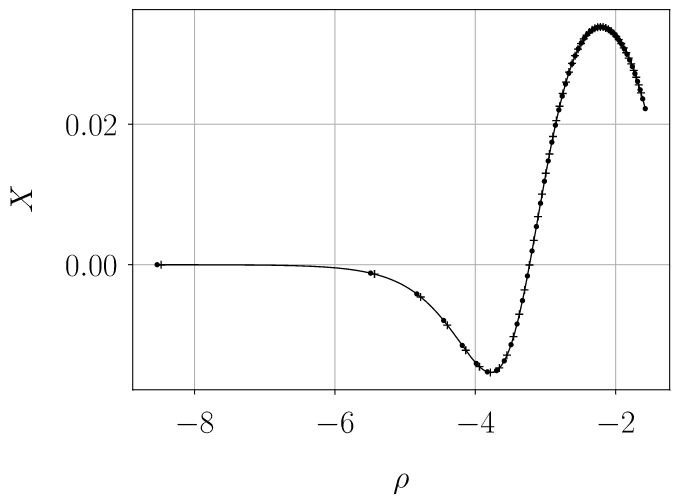}
  \caption{\label{f:universality}
    Universality:
    the scale-invariant variable $X^{(i)}(\tau,\rho)$ for a near-critical evolution of 
    the \emph{ingoing} family is plotted as a function of $\rho$
    at two different times $\tau = -1.8$ (left) and $\tau = -3.1$ (right) as a solid
    curve with dots at every tenth grid point.
    In the same plots, we also show $X^{(ii)}(\tau-\delta,\rho-\delta)$ for a near-critical
    evolution of the \emph{centred} family with plus symbols ($+$) at every tenth grid 
    point.
    The \emph{same} (family-dependent) constant offset $\delta = 0.245$ is used
    in both plots.
  }
\end{figure}

Finally we investigate if the critical solution is universal, i.e. independent of the
particular one-parameter family of initial data.
In Fig. \ref{f:universality} we again plot $X^{(i)}(\rho,\tau)$ for a near-critical 
evolution of the ingoing family as a function of $\rho$ at two different times $\tau$,
but this time we overlay $X^{(ii)}(\rho - \delta, \tau - \delta)$ for a near-critical evolution
of the \emph{centred} family, where $\delta$ is an overall family-dependent scale chosen 
such that the norm of the difference between the two solutions is minimal 
\emph{at one fixed} $\tau$. 
The fact that the curves nearly coincide also at a different time $\tau$ with the 
\emph{same} constant offset $\delta$ strongly supports the conjecture that the
critical solution is universal, as already argued in \cite{Choptuik1993}.

%%%%%%%%%%%%%%%%%%%%%%%%%%%%%%%%%%%%%%%%%%%%%%%%%%%%%%%%%%%%%%%%%%%%%%%%%%%%%%%
%%%%%%%%%%%%%%%%%%%%%%%%%%%%%%%%%%%%%%%%%%%%%%%%%%%%%%%%%%%%%%%%%%%%%%%%%%%%%%%

\section{Discussion}
\label{s:disc}

We presented a numerical method for gravitational collapse based on Cauchy 
evolution with an ingoing null boundary.
The method is similar in spirit to the excision method of Ripley \cite{Ripley2019}
but differs in that no grid points are removed from the computational domain;
rather, the grid remains fixed and only the coordinates are adapted along with the 
evolution.
This is achieved by adding a linear term to the shift vector that causes the coordinates
to ``zoom in'' isotropically towards the centre.
This linear term is a homogeneous solution to the isotropic spatial gauge condition.
Another important ingredient is the treatment of the lapse function.
We propose to use an advection equation for the lapse along the shift vector 
at the outer boundary in order to provide boundary values for the slicing condition 
(in our case, maximal slicing).
This corresponds to interpolating the lapse from the previous time step and 
minimises the amount of slice stretching as the lapse collapses towards zero
in the high curvature region in the centre.

We worked out the method in detail for the model problem of a massless scalar field
coupled to the Einstein equations in spherical symmetry.
Known results on critical behaviour \cite{Choptuik1993} are reproduced: 
the mass-scaling relation including its fine structure \cite{Hod1997,Gundlach1997}, 
the discrete self-similarity (echoing) of the critical solution and its universality 
among different families of initial data.
This demonstrates that the method is well suited to studying critical phenomena
in gravitational collapse, while being considerably simpler than more commonly
used methods that typically employ adaptive mesh refinement.

A price one has to pay for the simplicity of the method is that the outer boundary
radius $r_{\max}$ of the initial data slice needs to be chosen carefully so that a
sufficiently large region of spacetime where the evolution is close to the critical 
solution can be explored.
We optimised $r_{\max}$ using an outer bisection loop depending on the outcome of the
standard inner bisection along the critical parameter.
One might wonder if this makes the method overly computationally expensive.
Certainly in spherical symmetry this is not the case as a single evolution takes less than 
five minutes on a laptop even close to the critical point.
Furthermore one does not have to restart the bisection for the critical parameter
from the beginning for each $r_{\max}$ iteration; instead, a smaller interval
from the previous bisection can be used as an improved initial guess.
Whether the method is competitive in axisymmetry or without any spacetime symmetries
remains to be seen.

We have tried to alleviate the need for fine-tuning $r_{\max}$ by equipping the 
algorithm with a control system similar to the one described in \cite{Lindblom2007}:
make $\nu$ in \eref{e:bcbeta} time dependent and steer it so that a typical feature
of the solution such as the minimum of the outgoing expansion \eref{e:ahcondgen}
remains approximately at a constant coordinate radius.
The larger the value of $\nu$, the stronger the magnifying effect.
For this to work, $r_{\max}$ must be chosen somewhat larger than its optimal 
value for a null boundary, and $\nu$ must be taken somewhat larger than $1$ initially,
so that the control system has enough room to do its job.
While performing reasonably well at early times, we have found such a control system
to be ineffective in halting the rapid escape of the scalar field from the domain
that often occurs just before an apparent horizon forms if the initial $r_{\max}$ 
was chosen too small or the control system kept $\nu$ too large for too long a time. 
One should note that $\nu$ must not get smaller than $1$, otherwise the boundary
becomes timelike and boundary conditions for the evolved fields are needed.

Let us finally comment on other gauge conditions and less restrictive spacetime symmetries.
In axisymmetry there is the well-known quasi-isotropic (or isothermal) gauge in which
the spatial metric takes the form (compare \eref{e:metric})
\begin{equation}
  {}^{(3)} ds^2 = \phi^4 \rme^{2\eta/3} (\rmd r^2 + r^2 \rmd\theta^2 
  + \rme^{-2\eta} r^2 \sin^2\theta\,\rmd \varphi^2).
\end{equation}  
This has been used in much numerical work, including the first study of critical behaviour in vacuum axisymmetric gravitational collapse by Abrahams and Evans \cite{Abrahams1993},
as well as e.g. \cite{Rinne2008,Garfinkle2001,Choptuik2003,Choptuik2003a}).
The quasi-isotropic gauge condition admits homogeneous solutions 
analogous to the isotropic gauge condition in spherical symmetry, and our method
can be carried over with very few modifications.
Work along these lines is in progress.

It is conceivable that our method can be made to work with other classes of spatial
gauge conditions as well.
Any elliptic shift condition such as the minimal strain or minimal distortion conditions 
\cite{Smarr1978} requires boundary conditions, and the freedom in choosing the boundary 
data can be used to make the outer boundary an ingoing null surface.
Evolutionary shift conditions such as the hyperbolic Gamma-driver condition employed in 
some of the first successful binary black hole merger simulations \cite{Campanelli2006}
require initial conditions, and they could also be modified by adding lower-order terms,
which could be used to a similar effect.
These are interesting questions for further research.

Finally it should be stressed that this ingoing boundary method or the related 
method of Ripley \cite{Ripley2019} are not limited to studying critical collapse.
One can also start with a standard Cauchy evolution with timelike boundary 
(where of course boundary conditions must be imposed) and switch to the ingoing 
boundary method at a certain time.
Combinations with the outgoing boundary method of Bieri \etal \cite{Bieri2020}
are also possible.

%%%%%%%%%%%%%%%%%%%%%%%%%%%%%%%%%%%%%%%%%%%%%%%%%%%%%%%%%%%%%%%%%%%%%%%%%%%%%%%
%%%%%%%%%%%%%%%%%%%%%%%%%%%%%%%%%%%%%%%%%%%%%%%%%%%%%%%%%%%%%%%%%%%%%%%%%%%%%%%

\begin{acknowledgements}
  I am grateful to Ellery Ames for helpful comments on the manuscript.
\end{acknowledgements}  

%%%%%%%%%%%%%%%%%%%%%%%%%%%%%%%%%%%%%%%%%%%%%%%%%%%%%%%%%%%%%%%%%%%%%%%%%%%%%%%
%%%%%%%%%%%%%%%%%%%%%%%%%%%%%%%%%%%%%%%%%%%%%%%%%%%%%%%%%%%%%%%%%%%%%%%%%%%%%%%

%\bibliographystyle{orspphys}
%\bibliography{references}

\input{paper.bbl}\end{document}

%% file: fig1_labels.tex
\begin{picture}(0,0)%
\includegraphics{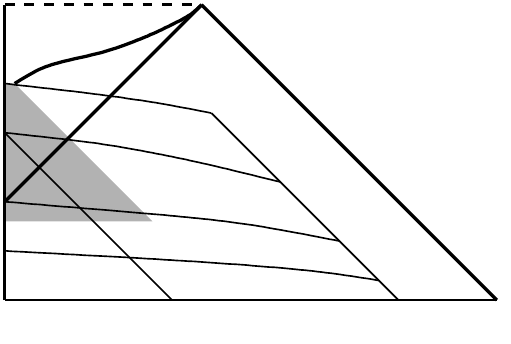}%
\end{picture}%
\setlength{\unitlength}{4144sp}%
\begingroup\makeatletter\ifx\SetFigFont\undefined%
\gdef\SetFigFont#1#2#3#4#5{%
  \reset@font\fontsize{#1}{#2pt}%
  \fontfamily{#3}\fontseries{#4}\fontshape{#5}%
  \selectfont}%
\fi\endgroup%
\begin{picture}(2332,1571)(204,-1610)
\put(316,-286){\makebox(0,0)[lb]{\smash{{\SetFigFont{9}{10.8}{\familydefault}{\mddefault}{\updefault}{\color[rgb]{0,0,0}AH}%
}}}}
\put(946,-376){\makebox(0,0)[lb]{\smash{{\SetFigFont{9}{10.8}{\familydefault}{\mddefault}{\updefault}{\color[rgb]{0,0,0}EH}%
}}}}
\put(1711,-601){\makebox(0,0)[lb]{\smash{{\SetFigFont{9}{10.8}{\familydefault}{\mddefault}{\updefault}{\color[rgb]{0,0,0}$\mathrsfs{I}^+$}%
}}}}
\put(2521,-1456){\makebox(0,0)[lb]{\smash{{\SetFigFont{9}{10.8}{\familydefault}{\mddefault}{\updefault}{\color[rgb]{0,0,0}$i^0$}%
}}}}
\put(901,-1546){\makebox(0,0)[lb]{\smash{{\SetFigFont{9}{10.8}{\familydefault}{\mddefault}{\updefault}{\color[rgb]{0,0,0}$r_1$}%
}}}}
\put(1351,-1546){\makebox(0,0)[lb]{\smash{{\SetFigFont{9}{10.8}{\familydefault}{\mddefault}{\updefault}{\color[rgb]{0,0,0}$t=0$}%
}}}}
\put(1981,-1546){\makebox(0,0)[lb]{\smash{{\SetFigFont{9}{10.8}{\familydefault}{\mddefault}{\updefault}{\color[rgb]{0,0,0}$r_2$}%
}}}}
\end{picture}%

%% file: paper.bbl
\begin{thebibliography}{10}
\providecommand{\url}[1]{{#1}}
\providecommand{\urlprefix}{URL }
\expandafter\ifx\csname urlstyle\endcsname\relax
  \providecommand{\doi}[1]{DOI \discretionary{}{}{}#1}\else
  \providecommand{\doi}{DOI \discretionary{}{}{}\begingroup
  \urlstyle{rm}\Url}\fi

\bibitem{Choptuik1993}
M.W. Choptuik: Universality and scaling in gravitational collapse of a massless
  scalar field. Phys.\ Rev.\ Lett. \textbf{70}, 9 (1993)

\bibitem{Gundlach2007}
C.~Gundlach, J.M. Mart\'in-Garc\'ia: Critical phenomena in gravitational
  collapse. Living Rev.\ Relativity \textbf{10}(5) (2007)

\bibitem{Berger1984}
M.J. Berger, J.~Oliger: Adaptive mesh refinement for hyperbolic partial
  differential equations. J.\ Comput.\ Phys. \textbf{53}, 484 (1984)

\bibitem{Hamade1996}
R.S. Hamad\'e, J.M. Stewart: The spherically symmetric collapse of a massless
  scalar field. Class.\ Quantum Grav. \textbf{13}, 497 (1996)

\bibitem{Garfinkle1995}
D.~Garfinkle: Choptuik scaling in null coordinates. Phys.\ Rev.\ D \textbf{51},
  5558 (1995)

\bibitem{Bieri2020}
L.~Bieri, D.~Garfinkle, S.T. Yau: A no-boundary method for numerical
  relativity. Class.\ Quantum Grav. \textbf{37}, 045015 (2020)

\bibitem{Sarbach2007}
O.~Sarbach: Absorbing boundary conditions for {E}instein's field equations.
  J.~Phys.: Conf.~Ser. \textbf{91}, 012005 (2007)

\bibitem{Winicour2012}
J.~Winicour: Characteristic evolution and matching. Living Rev.\ Relativity
  \textbf{15}(2) (2012)

\bibitem{Zenginoglu2008}
A.~Zengino\u{g}lu: Hyperbolodial evolution with the {E}instein equations.
  Class.\ Quantum Grav. \textbf{25}, 195025 (2008)

\bibitem{Moncrief2009}
V.~Moncrief, O.~Rinne: Regularity of the {E}instein equations at future null
  infinity. Class.\ Quantum Grav. \textbf{26}, 125010 (2009)

\bibitem{Rinne2010}
O.~Rinne: An axisymmetric evolution code for the {E}instein equations on
  hyperboloidal slices. Class.\ Quantum Grav. \textbf{27}, 035014 (2010)

\bibitem{Friedrich1983}
H.~Friedrich: Cauchy problems for the conformal vacuum field equations in
  general relativity. Commun.\ Math.\ Phys. \textbf{91}, 445 (1983)

\bibitem{Frauendiener2004}
J.~Frauendiener: Conformal infinity. Living Rev.\ Relativity \textbf{7}(1)
  (2004)

\bibitem{Ripley2019}
J.L. Ripley: Excision and avoiding the use of boundary conditions in numerical
  relativity. Class.\ Quantum Grav. \textbf{36}, 237001 (2020)

\bibitem{Arnowitt1962}
R.~Arnowitt, S.~Deser, C.W. Misner: The dynamics of general relativity. In
  \emph{Gravitation: an introduction to current research}, ed. by L.~Witten
  (Wiley, New York, 1962), chap.~7

\bibitem{Walsh2007}
D.M. Walsh: Non-uniqueness in conformal formulations of the {E}instein
  constraints. Class.\ Quantum Grav. \textbf{24}, 1911 (2007)

\bibitem{Rinne2008}
O.~Rinne: Constrained evolution in axisymmetry and the gravitational collapse
  of prolate {B}rill waves. Class.\ Quantum Grav. \textbf{25}, 135009 (2008)

\bibitem{Kreiss1973}
H.O. Kreiss, J.~Oliger: Methods for the approximate solution of time dependent
  problems.
\newblock Global Atmospheric Research Programme, Publication Series No. 10
  (1973)

\bibitem{Hod1997}
S.~Hod, T.~Piran: Fine structure of {C}hoptuik's mass-scaling relation. Phys.\
  Rev.\ D \textbf{55}, R440 (1997)

\bibitem{Gundlach1997}
C.~Gundlach: Understanding critical collapse of a scalar field. Phys.\ Rev.\ D
  \textbf{55}, 695 (1997)

\bibitem{Lindblom2007}
L.~Lindblom, K.D. Matthews, O.~Rinne, M.A. Scheel: Gauge drivers for the
  generalized harmonic {E}instein equations. Phys.\ Rev.\ D \textbf{77}, 084001
  (2008)

\bibitem{Abrahams1993}
A.M. Abrahams, C.R. Evans: Critical behavior and scaling in vacuum axisymmetric
  gravitational collapse. Phys.\ Rev.\ Lett. \textbf{70}, 2980 (1993)

\bibitem{Garfinkle2001}
D.~Garfinkle, G.C. Duncan: Numerical evolution of {B}rill waves. Phys.\ Rev.\ D
  \textbf{63}, 044011 (2001)

\bibitem{Choptuik2003}
M.W. Choptuik, E.W. Hirschmann, S.L. Liebling, F.~Pretorius: An axisymmetric
  gravitational collapse code. Class.\ Quantum Grav. \textbf{20}, 1857 (2003)

\bibitem{Choptuik2003a}
M.W. Choptuik, E.W. Hirschmann, S.L. Liebling, F.~Pretorius: Critical collapse
  of the massless scalar field in axisymmetry. Phys.\ Rev.\ D \textbf{68},
  044007 (2003)

\bibitem{Smarr1978}
L.~Smarr, J.W. York: Kinematical conditions in the construction of spacetime.
  Phys.\ Rev.\ D \textbf{17}, 2529 (1978)

\bibitem{Campanelli2006}
M.~Campanelli, C.O. Lousto, P.~Marronetti, Y.~Zlochower: Accurate evolutions of
  orbiting black-hole binaries without excision. Phys.\ Rev.\ Lett.
  \textbf{96}, 111101 (2006)

\end{thebibliography}
